\documentclass{kluwer}
\usepackage{epsf,amssymb}
\usepackage{graphics}
\usepackage{graphicx}
\usepackage{epsfig}

\def\eq#1{\begin{equation} {#1} \end{equation}}
\def\Nbar   {\hbox{$\bar N$}}
\def\Rbar   {\hbox{$\bar R$}}
\def\tautotbar   {\hbox{$\bar \tau_{tot}$}}
\def\Rav    {\hbox{$R_{\rm av}$}}
\def\Rs    {\hbox{$R_{\rm s}$}}

\newdisplay{guess}{Conjecture}

\begin{document}
\begin{article}

  \begin{opening}         

    \title{Spectral line and continuum  radiation propagation in a clumpy medium} 
    \author{John \surname{Conway}}
    \institute{Onsala Space Observatory, Sweden}
    \author{Moshe \surname{Elitzur}}
    \institute{Department of Physics and Astronomy, University of Kentucky, USA}
    \author{Rodrigo \surname{Parra}}
    \institute{Onsala Space Observatory, Sweden}
    \runningauthor{Conway, Elitzur and Parra}
    \runningtitle{Spectral lines in a clumpy medium}
    \date{April 15, 2004}

    \begin{abstract}
     We discuss the propagation of spectral line and continuum  
     radiation in a clumpy  medium and give general expressions for 
     the observed absorption or emission  from a cloud population. 
     We show that the affect of the 
     medium  clumpiness can usually be characterised by a single number
     multiplying the mean column opacity.
     Our result provides a simpler proof and generalization 
     of the result of  Martin et al. (1984). The formalism  
     provides  a simple way to understand the effects of clumping on 
     molecular line profiles and ratios; for example how clumping effects the 
     interpretation of $^{13}$CO(1-0) to $^{12}$CO(1-0) line ratios. It also 
     can be used as a propagation operator in physical models of clumpy 
     media where the incident radiation effects the spectral line 
     emissivity. We  are working to extend the formalism to the propagation 
     of masers in 
     a clumpy medium, but in this case there are special difficulties 
     because formal expectation values are not characteristic of observations
     because they are biased by rare events. 

    \end{abstract}

  \end{opening}

\section{Introduction}

The problem of the propagation of radiation in a non-uniform 
medium  is a general one in astrophysics, which arises in many 
different contexts. We consider here the  common case that the 
medium consists of clumps or clouds with a low volume filling 
factor, but potentially  high covering factor.  
These clumps  or clouds may possess a range of properties. 
This paper gives a summary of our derivation and results, 
a more detailed proof will be published elsewhere 
(Conway, 
Elitzur and Parra 2004, in prep). 

The specific problem we address is the expected  response of a
telescope beam observing a low filling factor  clumpy region where
the beamwidth is much larger than the typical cloud diameter. In this 
case we can show that that expectation value of the beam average equals
the ensemble expectation value over many realisations of observations
along the line of sight (LOS).
%Different lines of sight will have different numbers of clouds. In this situation it is important 
% to realise that the mean response will not in general be the same as the response of having the mean number of clouds along every LOS.
% Instead the statistical variations between different lines of sight
% must be fully taken into account.
We first (see Section 2) analyse the case  of continuum (e.g. dust) clouds 
absorbing background radiation, then spectral  line absorption (Section 3). In Section 4 we consider spectral line emission and apply the formalism to the interpretation of CO line ratios. Finally in Section 5 we draw conclusions and describe future work.

\section{Continuum Absorption Radiation Transfer}

Let us consider continuum radiation transfer in a clumpy medium of
dust clouds. First assume that all clouds/clumps are identical. Assume 
further that radiation $I$ entering a clump emerges as  $I\cdot R$, where the clump response is $R$. In terms of cloud opacity $R=e^{-\tau}$. Hitting successively $k$ clumps that all have the same response factor produces the aggregate response $R^k$. Implicit in this assumption is that the interaction with the radiation does not change the clump response factor $R$. 
It can be shown that provided the clump volume filling factor is small the probability of getting  $k$  clumps along a given line of sight always obeys a Poisson distribution, characterised by one number, the mean $\Nbar$ along a LOS. In this case
of one clump type it can be shown that the mean response is 
\eq{
    \bar R= e^{-t}, \qquad t = \Nbar(1 - e^{-\tau})
}
as previously derived by Natta \&  Panagia (1992).

Next consider the case of many types of clumps with response factors 
$R_i$ (these could be clump types having different optical depths) each obeying Poisson statistics. If the fraction of clumps of each type is $f_{i}$ then
(see Conway,  Elitzur and Parra 2004, in prep)
\eq{
    \Rbar = \exp\left(\Nbar[\Rav - 1]\right),
    \qquad \hbox{where}\quad
    \Rav = \sum_i f_i R_i
}

\noindent This means that the problem of 
a mixture of cloud optical depths ($\tau_{i}$) is equivalent to that of the single cloud type having an average $\tau_{\rm av}$ obtained from
\eq{
  \Rav =   e^{-\tau_{\rm av}} = \sum_i f_i e^{-\tau_{i}}
}  

We can generalize the result in Eq. (2) to the case where the clump type varies continuously. For instance if the clump type is described 
by continuous variables $t$ and $r$ (which may correspond for instance
to peak optical depth and distance from the centre of clouds which 
have spatial structure) then we have

\eq{
    \Rav = \int \int R(r,t)\eta(r,t)drdt = \int \int e^{-\tau(r,t)} \eta(r,t)drdt
}

\noindent and likewise for any number of dimensions of continuous parameters, where $\eta$ is the distribution function of clump 
types.

\section{Line Radiation Transfer - Absorption}

In the spectral line absorption problem we aim to calculate the expectation value of a frequency 
dependent absorption response  function 
given a population of clouds.  This cloud population has in general
 a distribution of central LOS velocities and  a range of  
spectral profiles and spatial structures. Observed absorption lines are 
generally published as a function of Doppler velocity 
$v$, the response at velocity $R(v)$  being understood to represent 
the response at observing frequency $\nu = \nu_0(1 + v/c)$ where $\nu_o$ is 
the line's rest frequency.  Let us concentrate on calculating the response
at a particular velocity $v$, $R(v)$. In principle clouds having  all possible centre 
velocities $v'$ contribute to the spectral response at velocity $v$. The centre velocity  of each cloud $v'$ can simply be thought of as yet another  continuous variable describing the cloud properties, so the full distribution 
function becomes $\eta(r,t,v')$.
For spectral line clouds the response function is $R_{v}(r,t,v')$ which if  
cloud types are defined to have fixed spectral profiles which don't vary 
with centre velocity  can be written $R(r,t,v-v')$. Hence applying
Eq. (2) and (4) we obtain

\eq{\Rbar(v) = \exp\left(\Nbar[\Rav(v) - 1]\right)
} 

\eq{\hbox{where}\quad     \Rav(v) = \int \int \int R(r,t,v-v')\eta(r,t,v') dv' dr dt
}

\noindent As before $\Nbar$ is the mean number of clouds (at any velocity) 
per LOS. $\Rav(v)$ can be interpreted at each $r,t$ value as a 
convolution of the cloud  velocity profile for this $r$,$t$, $R(r,t,v)$  
with the velocity distribution function  of the cloud population 
$\eta(r,t,v)$  followed by an integration over  all $r$,$t$ (and any 
other continuous cloud parameters).

If the fractional distribution of clouds of different kinds is
independent of velocity, then $\eta(r,t,v) = f(r,t) q(v)$ is a separable  function. In this case we can define a mean cloud
spectral response

\eq{ 
    \Rs(v) = \int \int R(r,t,v) f(r,t) dr dt
}

\noindent In this separable case the expectation value of 
the emerging spectrum is 
\eq{ 
\Rbar(v) = \exp\left(-\lambda(v) * C(v) \right)
}\noindent In this expression $*$ denotes convolution, 
$\lambda(v) = \Nbar q(v)$ describes the cloud velocity distribution;
more precisely defined as the mean number of clouds per 
unit velocity per LOS. The quantity  
$C(v) = 1-\Rs(v)$ = \\ $<1 -e^{-\tau(v)}>$,  
where $\tau(v)$ is the opacity profile of each clump type, and the 
brackets denote an average over the clump population.
The $C(v)$ can  be interpreted  as the mean cloud emission 
profile. 
In the normal case that the $\lambda(v)$ distribution is much 
broader in velocity than $C(v)$ we obtain the good approximation that 
$\Rbar(v) \approx  \exp\left(-C_{A}\lambda(v) \right)$
where $C_{A} = \int C(v) dv$ has dimensions of velocity and 
can be considered the equivalent width  of the mean cloud 
profile. 

Another way to view the effect of clumping is to note that 
\eq{ 
\Rbar(v) \approx  \exp\left(-K\tautotbar(v) \right)
}
\noindent where $\tautotbar(v) = \lambda(v) * < \tau(v) >$ 
is the mean line of sight opacity and $K$ is a factor taking into account
clumping which is 

\eq{ 
 K =   { \int < 1 -e^{-\tau(v)}> dv   \over 
         \int <\tau(v)>dv}
} 

\noindent Remarkably we find that the effect of an arbitrarily 
complex  clump distribution can usually  be reduced to a single
number, $K$. This factor $K$ reduces the effective opacity of 
a clumpy medium compared to a smooth gas of the same mean column 
density. Martin et al. (1984) partially derived our results, but 
their derivation  was less direct and the full generality of the 
result was not  stated.

\section{Emission Line Spectra}

The expressions for opacity derived above can be adapted 
(Conway, Elitzur and Parra 2004, in prep) to predict the 
emissivity in the case of a population of clouds 
with  a mixture of excitation temperatures and statistical
 properties  along the line of sight. Here
we discuss only the simplest case that all clouds have the same excitation
temperature $T_{e}$. In this case the observed spectral brightness temperature is

\eq{ T(v) = T_{e} [ 1- \Rbar(v)] = T_{e}[1- \exp\left(-K\tautotbar(v) \right)]
} 

To illustrate the impact of clumping on observed line profiles and 
line ratios we show an example in Fig 1 involving the 
$^{12}$CO(1-0)and $^{13}$CO(1-0) lines. In this example the cloud 
population  consists of only one clump type with a Gaussian opacity  
velocity profile $\tau(v)$, and a much broader Gaussian $\lambda(v)$ 
cloud velocity distribution. The expectation value of the total 
opacity along each LOS $\tautotbar(v)=\lambda(v)*\tau(v)$   is 
therefore also a Gaussian whose peak we term the  'peak total 
opacity'.

The different panels for Fig 1 have different combinations 
of the $^{12}$CO(1-0) peak total and peak cloud opacities. The ratio 
of intrinsic $^{12}$CO(1-0) and $^{13}$CO(1-0) opacities is set to 
$60$ so that in all the panels of Fig 1, the $^{13}$CO(1-0) total and
cloud opacities are $<<1$ and hence are optically thin. In all cases 
therefore the $^{13}$CO(1-0) line profiles are Gaussian with peak 
value (in units of $T_{e}$) equal to the total $^{12}$CO(1-0) opacity 
divided by 60. In contrast the $^{12}$CO(1-0) line profiles 
and their ratio to the $^{13}$CO(1-0) lines depend critically on 
the clump opacity which affects $K$. Only for the case in the top left 
panel, which has 
 both a low total and low clump opacity, is the observed line ratio equal to the intrinsic value. If either or both total and clump 
opacities are larger than 1 then a smaller line ratio is observed, 
but it is impossible just from the line ratio at line centre
to distinguish the different cases.
%  Note that a flat-topped   
% $^{12}$O(1-0) line profile (see Fig 1, top right) when  present 
% does indicate a large total opacity but the absence of such a 
% profile (see Fig 1, bottom right) does not rule out a large total
% opacity. 

An important consequence of Eq. (9) is that when the cloud
velocity distribution is much broader than individual clouds it 
is in principle impossible from the $^{12}$CO(1-0) and $^{13}$CO(1-0)
spectra alone to separate the effects of clumping and
variations in the intrinsic $^{12}$CO(1-0)/$^{13}$CO(1-0) opacity 
ratio. In going from  panels on the top row with low opacity 
clumps to the bottom row with large  clumps the $^{12}$CO(1-0) 
clump factor $K$ is reduced. The resulting spectra are the same 
as if there was no clumping but the intrinsic 
$^{12}$CO(1-0)/$^{13}$CO(1-0) opacity ratio was reduced.

\section{Conclusions and Future Work}

We have derived general expressions for both continuum and
spectral line radiation propagation
in a clumpy medium. These expressions hold provided 
only that the clumps have a small volume filling factor and that the local emissivity is not itself affected by the local line 
spectra. Eq. (8),(9) and (10) provide simple expressions that can be used by observers to interpret observed line profiles. In Section 4
we illustrated the application of this equation to CO line 
emission profiles from a clumpy medium. Although the results 
presented in Section 4 are not new our equations provide a way 
to understand line profiles from clumpy media in a systematic 
and quantitative way.

We are working to  extend our formalism to predict variances
and higher order statistics of spectra from clumpy media. We 
also hope to use  our formalism as a propagator
term  in physical  spectral line models of clumpy media. 
In general in such models emissivity will depend on the 
local spectrum so such models must be iterative. Finally 
we are actively working (see Parra et al. these proceedings)
 on the problem of maser propagation
in a clumpy medium. Formally exactly the same equations apply 
as in the thermal case, the only difference is the use of 
negative opacities. However as well as physical  complications of 
maser saturation there is a more subtle effect that even for unsaturated 
masers the formal means predicted by the equations are not good 
predictors of observations. The reason for this is 
that these mean values  are dominated by events causing 
very large amplification which are also very improbable, 
such that they never occur within a typical source.

 %%%%%%%%%%%%%%%%%%%%%%%%%%%%%%%%%%%%%%%%%%%%%%%%%%%%%%%%%%%%%
\begin{figure}%[ht!]
\begin{center}
$\begin{array}{c@{\hspace{1cm}}c}
%\textrm{EVN+MERLIN obsevations}&\textrm{Model}\\
\includegraphics[width=0.7\columnwidth]{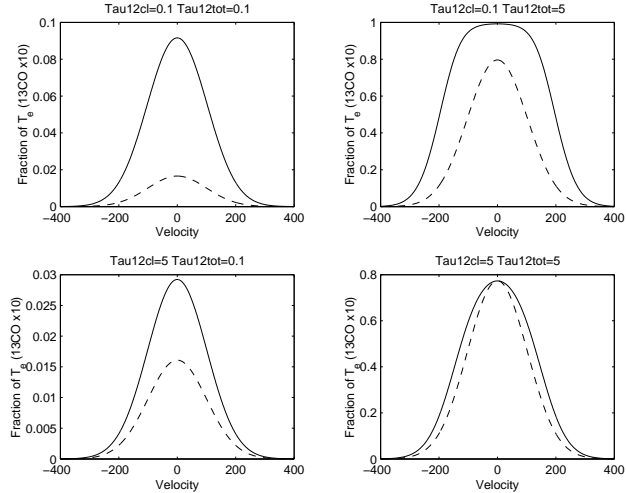} 
\end{array}$
\end{center}
\caption{Example $^{12}$CO(1-0) and $^{13}$CO(1-0) line profiles (shown respectively by 
solid and dashed lines) for different combinations of cloud and mean column opacities. 
To make them more visible the $^{13}$CO(1-0)
profiles have been multiplied by 10. 
The intrinsic ratio of opacity in the two transitions is assumed to be 60.
Both the internal cloud opacity profiles of the clouds in the two transitions and the
cloud velocity distribution are  assumed to be Gaussian with the latter 
having a FWHM 10 times larger. The top row shows  the case of low opacity 
in each $^{12}$CO(1-0) cloud, and a varying mean {\it total} $^{12}$CO(1-0)
opacity. 
The bottom row shows the case of high opacity $^{12}$CO(1-0)
clouds. Note the different vertical scales of each plot.
Only when the total and cloud $^{12}$CO(1-0) opacities are both less
than one is the observed $^{12}$CO(1-0) to $^{13}$CO(1-0) ratio close to 
the intrinsic  value of 60.}
\label{fi:figure1}
\end{figure}

\end{article}

\begin{thebibliography}{}


\bibitem[1984]{martin84}
Martin H.M., Hills R.E., \& Sanders D.B., 1984, MNRAS, 208, 35


\bibitem[1992]{natta92}
Natta A. \&  Panagia N., 1984, ApJ, 287, 228


\end{thebibliography}
\end{document}